\documentclass[aps,nofootinbib,pra,notitlepage,superscriptaddress,10pt,twocolumn]{revtex4-2}
\usepackage{amsfonts,amsmath,amssymb,amsthm,revsymb4-1}
\usepackage{array,bm,color,upgreek,url}
\usepackage[pdfusetitle,colorlinks=true]{hyperref}
\usepackage{epsfig,graphicx}
\graphicspath{{./figures/}}

\makeatletter
\DeclareFontFamily{OMX}{MnSymbolE}{}
\DeclareSymbolFont{MnLargeSymbols}{OMX}{MnSymbolE}{m}{n}
\SetSymbolFont{MnLargeSymbols}{bold}{OMX}{MnSymbolE}{b}{n}
\DeclareFontShape{OMX}{MnSymbolE}{m}{n}{
    <-6>  MnSymbolE5
   <6-7>  MnSymbolE6
   <7-8>  MnSymbolE7
   <8-9>  MnSymbolE8
   <9-10> MnSymbolE9
  <10-12> MnSymbolE10
  <12->   MnSymbolE12
}{}
\DeclareFontShape{OMX}{MnSymbolE}{b}{n}{
    <-6>  MnSymbolE-Bold5
   <6-7>  MnSymbolE-Bold6
   <7-8>  MnSymbolE-Bold7
   <8-9>  MnSymbolE-Bold8
   <9-10> MnSymbolE-Bold9
  <10-12> MnSymbolE-Bold10
  <12->   MnSymbolE-Bold12
}{}

\let\llangle\@undefined
\let\rrangle\@undefined
\DeclareMathDelimiter{\llangle}{\mathopen}%
                     {MnLargeSymbols}{'164}{MnLargeSymbols}{'164}
\DeclareMathDelimiter{\rrangle}{\mathclose}%
                     {MnLargeSymbols}{'171}{MnLargeSymbols}{'171}
\makeatother
\usepackage{txfonts}

\begin{document}

\newcommand{\cE}{\mathcal{E}}
\newcommand{\cL}{\mathcal{L}}
\newcommand{\cB}{\mathcal{B}}
\newcommand{\cH}{\mathcal{H}}

\newcommand{\reals}{\mathbb{R}}
\newcommand{\complex}{\mathbb{R}}

\newcommand{\expect}[1]{\ensuremath{\left\langle#1\right\rangle}}
\newcommand{\ket}[1]{\ensuremath{\left|#1\right\rangle}}
\newcommand{\bra}[1]{\ensuremath{\left\langle#1\right|}}
\newcommand{\braket}[2]{\ensuremath{\left\langle#1|#2\right\rangle}}
\newcommand{\ketbra}[2]{\ket{#1}\!\!\bra{#2}}
\newcommand{\braopket}[3]{\ensuremath{\bra{#1}#2\ket{#3}}}
\newcommand{\proj}[1]{\ketbra{#1}{#1}}
\newcommand{\sket}[1]{\ensuremath{\left|#1\right\rrangle}}
\newcommand{\sbra}[1]{\ensuremath{\left\llangle#1\right|}}
\newcommand{\sbraket}[2]{\ensuremath{\left\llangle#1|#2\right\rrangle}}
\newcommand{\sketbra}[2]{\sket{#1}\!\!\sbra{#2}}
\newcommand{\sbraopket}[3]{\ensuremath{\sbra{#1}#2\sket{#3}}}
\newcommand{\sproj}[1]{\sketbra{#1}{#1}}

\def\Id{1\!\mathrm{l}}
\newcommand{\Tr}{\mathrm{Tr}}
\newcommand{\order}[1]{\mathcal{O}\left( #1 \right)}
\newcommand{\trans}[1]{#1^\mathsf{T}}
\newcommand{\T}{\mathsf{T}}
\newcommand{\erf}[1]{Eq.~(\ref{#1})}

\title{Quantum Characterization, Verification, and Validation}

\author{Robin Blume-Kohout}
\author{Timothy Proctor}
\author{Kevin Young}
\affiliation{Quantum Performance Laboratory, Sandia National Laboratories, Albuquerque, NM 87185 and Livermore, CA 94550}
\date{March 20, 2025}

\begin{abstract}
Quantum characterization, verification, and validation (QCVV) is a set of techniques to probe, describe, and assess the behavior of quantum bits (qubits), quantum information-processing registers, and quantum computers.  QCVV protocols probe and describe the effects of unwanted decoherence so that it can be eliminated or mitigated.  They can be usefully divided into \emph{characterization} techniques that estimate predictive models for a device's behavior from data, and \emph{benchmarking} techniques that assess overall performance of a device.  In this introductory article, we briefly summarize the history of QCVV, introduce the mathematical models and metrics upon which it relies, and then summarize the foundational fields of tomography, randomized benchmarking, and holistic benchmarks.  We conclude with brief descriptions of (and references to) advanced topics including gate set tomography, phase estimation, Pauli noise learning, characterization of mid-circuit measurements and non-Markovianity, classical shadows, verification and certification, and logical qubit assessment.
\end{abstract}

\maketitle

%~+~+~+~+~+~+~+~+~+~ Body ~+~+~+~+~+~+~+~+~+~
% ==============================================================================
% Section: Introduction
% ==============================================================================
\section{Executive summary}

\textbf{Quantum characterization, verification, and validation} (QCVV) is a set of techniques to probe, describe, and assess the behavior of quantum bits (qubits), quantum information-processing registers, and quantum computers.  These devices can enable remarkable information-processing algorithms if they are protected from noisy environment-induced decoherence.  QCVV protocols probe and describe the effects of unwanted decoherence so that it can be eliminated or mitigated.  QCVV is surprisingly unrelated to ``verification and validation'' (V\&V) in engineering.  Whereas V\&V tends to be prescriptive, QCVV is largely descriptive and seeks to answer ``What's happening inside the quantum device?''  QCVV methods can be usefully divided into \emph{characterization} techniques that estimate predictive models for a device's behavior from data, and \emph{benchmarking} techniques that assess overall performance of a device.  QCVV relies upon a standard set of mathematical models for noisy quantum systems, the \emph{open quantum systems} model, and on a growing set of widely used performance metrics including fidelities and statistical distances.

The earliest QCVV protocols were \emph{tomography} protocols that used the results of many measurements on identically prepared copies of a quantum system to reconstruct a mathematical model of its state or dynamical behavior.  Tomography actually predates quantum computing, and flourished first in the field of quantum optics.  Although tomographic protocols were adopted by early qubit engineers and scientists, and are used to this day, it rapidly became obvious that tomography does not scale well to many-qubit systems because the standard mathematical models for their behavior grow exponentially in size and complexity with the number of qubits.  Although innovative ``ansatz'' techniques have been used to design and tomograph smaller models, these scalability challenges spurred alternatives to tomography.

The most prominent alternative is \emph{randomized benchmarking} (RB).  RB is fundamentally a technique for mixing or ``twirling'' all errors in a qubit or quantum computer into a single error rate that can be used as a benchmark.  It provides much less information than tomography, but at significantly lower cost.  RB is ubiquitous as a simple, effective technique for quantifying the performance of few-qubit systems.  In recent years, truly scalable RB techniques have been created that can efficiently benchmark registers with tens or hundreds of qubits.  A semi-scalable variant of RB, cross-entropy benchmarking (XEB) was intrinsic to Google Quantum AI's demonstration of ``quantum supremacy'' in 2019 \cite{arute2019quantum}.  Many variants of RB have been proposed and deployed for specific purposes.

More recently, \emph{holistic benchmarks} that assess the performance of integrated quantum computers have been invented and deployed.  One of the first, and arguably still the most prominent, is quantum volume.  Holistic benchmarks attempt to simulate ``real-world'' workflows or use cases, sometimes by running tiny instances of quantum algorithms, or by executing random circuits with interesting or useful properties.  Like RB, holistic benchmarks tend to measure effective error rates, but unlike RB they are often intended to exercise all or many levels of the quantum computer's execution stack, such as (e.g.) compilation and routing.

A wide range of more specialized and advanced QCVV techniques have been invented and deployed, often building on the foundation blocks of tomography, RB, and holistic benchmarking.  These include gate set tomography (GST), phase (eigenvalue) estimation techniques such as robust phase estimation (RPE), Pauli noise learning and cycle error reconstruction, techniques focused on assessing mid-circuit measurements (MCMs), protocols that probe and characterize non-Markovian errors (notably \emph{noise spectroscopy}), classical shadow techniques based on randomized measurements, and verification or certification approaches that attempt to confirm correct functioning of a quantum device.  As quantum computing progresses toward fault tolerant utility-scale machines, QCVV of \emph{logical qubits} that use quantum error correction to protect a small high-fidelity quantum register within an array of noisy physical qubits is growing rapidly in relevance and interest.

\textbf{Keywords:} quantum characterization, verification, and validation; quantum computers; quantum benchmarking; quantum tomography; randomized benchmarking; quantum volume; quantum characterization; decoherence; fidelity
\newpage

\section{Introduction}
Quantum characterization, verification, and validation (QCVV) is a set of techniques to probe, describe, and assess the behavior of quantum information processing devices.  A scientific subfield of quantum information science, it is primarily (but not exclusively) concerned with qubits, quantum logic operations, and quantum processors that are designed for quantum computing.

Physical systems and devices that obey the laws of quantum theory but not those of classical physics can perform information-processing tasks that are interesting, counterintuitive, and potentially useful \cite{NielsenBook2000}.  These tasks include secure communication, ultra-sensitive metrology, and fast computation of chemical energies and large integers' factors.  The devices that perform these quantum information-processing tasks must be extremely small -- e.g. single atoms, electrons, or photons -- and/or kept extremely cold at millikelvin temperatures in a dilution refrigerator, because quantum information is exceptionally fragile.  Observing or measuring a quantum information-processing device -- whether by a human, or just by its environment -- changes its state and damages the quantum information held in it, through a process called \textit{decoherence} \cite{zurek2003decoherence}.  Scientists and engineers can't simply look inside a device and watch it run, since this would disrupt the very behavior they want to study.  QCVV allows them to reconstruct a device's behavior from indirect measurements.

This article will briefly introduce the most important aspects of QCVV, and the most prominent techniques and protocols.  Describing (or even mentioning) every key idea is impossible in an article of this length.  This article focuses on QCVV as applied to quantum bits or \textit{qubits} designed to implement logic gates for quantum computing, at the expense of non-qubit systems, non-gate-model computing, and communication or metrology.  It is intended as an introduction for anyone who wants to understand how real-world quantum information processors are probed and assessed, and a gateway to the scientific literature.

\section{History}

The term ``QCVV'' is believed to have emerged in the early 2010s, in US Government workshops and research funding announcements, to describe a research area that has surprisingly little in common with verification and validation (``V\&V''). The latter are well-defined engineering concepts often described as ``Are you building the right thing?'' (validation) and  ``Are you building it right?'' (verification).  QCVV is primarily concerned with characterization and benchmarking of qubits and other quantum systems that hold or process quantum information.  Its development has consistently been driven by the emergence of new quantum hardware.

Quantum theory, and the widespread recognition that tiny physical systems don't obey the predictions of classical physics, originated in the 1920s.  Quantum information science emerged in the 1980s, as scientists like Feynman \cite{feynman1986quantum}, Bennett \cite{bennett1984proceedings}, and Deutsch \cite{deutsch1985quantum} began to develop the notion that quantum systems might store and process information differently.  In the half-century between 1930 and 1980, techniques for probing and understanding the behavior of quantum systems were pioneered under the general name of \textit{spectroscopy} by physicists, chemists, and (especially) practitioners of nuclear magnetic resonance (NMR).  Many of these techniques, such as Rabi and Ramsey oscillations and measurements of $T_1$ and $T_2$ decay times, are used today as ad hoc QCVV protocols.

True QCVV protocols probe \textit{individual} quantum systems (rather than ensembles), and measure properties that are important for information processing (rather than for physics or chemistry).  \textit{Quantum state tomography}, the earliest example, dates to 1989 \cite{Royer1989-kp,Vogel1989-vt,Wootters1989-kq} (although related ideas can be traced to work from the 1960s \cite{Gale1968-mk}).  State tomography combines the results of measuring many different and incompatible observables on independent (but identically prepared) samples of a quantum system, to deduce its quantum state.  \textit{Quantum process tomography} was introduced in 1997 \cite{Chuang1997-vf} to do the same reconstruction for quantum dynamical operations such as computational logic gates \cite{Poyatos1997-mz}.

The explosion of usable qubit systems in the 2000s revealed two significant limitations of state and process tomography:  they require a lot of experimental data, and conclusions drawn from them can be rendered unreliable by \textit{state preparation and measurement} (SPAM) errors.  These limitations motivated the invention of \textit{randomized benchmarking} (RB) \cite{Emerson2005-sd,Knill2008-qe,Magesan2011-ra}, the first ``modern'' QCVV protocol that accounts for SPAM errors and measures an objective quality metric, circa 2005.  The original RB protocol measures and reports a single averaged error rate for an entire quantum processor.  It spawned a large family of derivative protocols with the same highly focused philosophy of isolating one useful ``error rate'' at a time.  Around 2012, the invention of \textit{gate set tomography} \cite{Merkel2013-me,Blume-Kohout2013-mf,Greenbaum2015-tr} merged the strengths of RB (invariance to SPAM) and tomography (detailed predictive characterization of individual operations) into a single protocol for building detailed predictive models of logic gates.

Around 2015, the appearance of integrated processors with 5 or more qubits \cite{Linke2017-ht} spurred development of \textit{holistic benchmarking} protocols that could be used to assess and compare them.  Small instances of quantum algorithms had been run on multiqubit NMR systems as early as 2001 \cite{vandersypen2001experimental}, and these demonstrations were repeated in the 2010s with much higher fidelity \cite{Lucero2012-kd,Linke2017-ht,Omalley2016-as}.  More sophisticated  benchmarks such as quantum volume \cite{Cross2019-el}, cross-entropy benchmarking \cite{Boixo2018-wy}, and scalable RB protocols \cite{Erhard2019-ig,Proctor2019-ma,Proctor2022-nt} were developed to systematically capture important features of quantum computer performance.

The latest evolution of QCVV (as of 2025) is the assessment of \textit{logical} qubit systems that encode a few qubits deeply into a large multipartite quantum system (e.g., many physical qubits) and protect them using active quantum error correction (QEC).  The first logical qubit systems were demonstrated beginning in 2016 \cite{Ofek2016-ux,Egan2021-sw,Ryan-Anderson2021-xv,Krinner2022-ff,Postler2022-ee,Abobeih2022-ed,Google2023-si,Gupta2023-nb,Bluvstein2023-my}, with performance beyond the threshold for useful reduction of error rates demonstrated in 2024 \cite{Google2025-xr}.  If quantum computing technology continues to progress, logical-qubit QCVV will necessarily evolve and spawn new protocols for characterizing large, high-fidelity quantum processors that execute increasingly complex subroutines of utility-scale quantum algorithms, and benchmarking their growth toward quantum utility \cite{proctor2025benchmarking}.

\section{Models and metrics}

QCVV's purpose is to understand and predict the behavior of real-world quantum information processing devices.  To do so, it relies heavily on mathematical models of quantum physical systems.  The heart of a quantum processor is an information-carrying register, often comprised of $n$ ``physical qubit'' systems that can each hold a single quantum bit of information.  Accordingly, models for such registers lie at the heart of QCVV.  Other parts of the quantum device, such as cooling and control hardware, are only of interest to QCVV protocols inasmuch as they impact the performance or behavior of the quantum data register.  Models of quantum data registers provide mathematical descriptions of three key components: quantum \textit{states}, control \textit{operations}, and readout \textit{measurements} \cite{NielsenBook2000,Kraus1983-gg}.

Scientists describe an \textit{ideal} quantum register using a $d$-dimensional complex vector space $\complex^d$, called a \textit{Hilbert space}, where $d$ is the number of perfectly distinguishable states in which the register could be found.  A single qubit has $d=2$, and $n$ qubits have $d=2^N$.  The register's \textit{state} is a normalized vector $\ket\psi$ in its Hilbert space.  When the state is changed by some control operation, that change is described by a $d\times d$ complex matrix $U$ that is unitary ($U^\dagger U = \Id$), so $\ket{\psi}_{\mathrm{initial}} \to \ket{\psi}_{\mathrm{final}} = U\ket{\psi}_{\mathrm{initial}}$.  Reading out the register's value is called \textit{measurement}.  A quantum register can be measured in many inequivalent ways.  Each measurement is described by some orthonormal basis of $d$ vectors $\{\bra{\phi_k}:\ k=1\ldots d\}$, and the probability of seeing the $k$\textsuperscript{th} outcome is $\mathrm{Pr}(k|\psi) = \left|\braket{\phi_k}{\psi}\right|^2$ (this is called \textit{Born's rule}).

Because QCVV is concerned with flaws and errors, it requires a more complex model that can describe \textit{noisy} registers, which physicists call \textit{open quantum systems}.  In the open systems model, a register's state is described by a $d\times d$ \textit{density matrix} $\rho$ that (1) is Hermitian ($\rho^\dagger = \rho$), (2) has non-negative eigenvalues ($\rho \geq 0$), and (3) has trace 1 ($\Tr[\rho]=1$).  Any state that can be described by a state vector $\ket{\psi}$ is called \textit{pure}, and the corresponding density matrix is the rank-1 projector $\proj{\psi}$.  States of noisy registers can only be represented by density matrices with rank greater than 1, and are called \textit{mixed}.  Ideal operations are still described by unitary matrices, which act on mixed states as $\rho_{\mathrm{initial}} \to \rho_{\mathrm{final}} = U\rho_{\mathrm{initial}}U^\dagger$.  \textit{Noisy} operations are described by \textit{completely positive trace-preserving (CPTP) linear maps} on density matrices, also known as \textit{quantum channels} or \textit{superoperators}, which can always be written in an operator-sum representation as
\begin{equation}
    \rho_{\mathrm{initial}} \to \rho_{\mathrm{final}} = \sum_k{ K_k\rho_{\mathrm{initial}}K_k^\dagger},
\end{equation}
where $K_k$ is a $d\times d$ matrix and $\sum_k{K_k^\dagger K_k} = \Id$.  A noisy readout measurement with $m$ possible outcomes is described by a set of $m$ positive operators $\{E_k:\ k=1\ldots m\}$, known as a \textit{positive operator-valued measure} (POVM), where Born's rule for measurement probabilities becomes $\mathrm{Pr}(k|\psi) = \Tr[ E_k\rho ]$.  

Most QCVV theories and techniques assume that information-processing operations on quantum registers are well-described by this model.  For example, in a quantum computer, each physical qubit is prepared in some state (density matrix) $\rho$, and then a computation is executed by performing logic gates (CPTP maps) $G_1, G_2, \ldots$, and finally the answer is read out by performing a measurement (POVM) $\{E_k\}$.  A scientist could model this computation by computing the probability of each possible outcome $k$ as a product of matrices.  However, if anything external to the register acts as a ``memory'' and mediates correlations between times (e.g.~between $G_1$ and $G_2$), it can (and does) produce \textit{non-Markovian} dynamics \cite{shrikant2023quantum} that deviate from this model and may invalidate QCVV protocols that rely on it.

The primary task of QCVV is estimating these models and/or properties of them from experimental data.  Density matrices, CPTP maps, and POVMs are complex and unwieldy, especially for large registers.  So a variety of \textit{metrics} \cite{Gilchrist2005-pf} are used to summarize the quality of a state, operation, measurement, or entire system. These metrics can be computed from a fully- or partially-estimated model, or in some cases a metric can be estimated directly from data without learning very much else about the model.  Some popular metrics are defined by a particular protocol, essentially as the quantity measured by that protocol.  Others capture fundamental aspects of information-processing performance.  The most commonly encountered metrics are \textit{fidelities} \cite{Jozsa1994-xu} and \textit{statistical distances} \cite{Helstrom1969-tq}.

Fidelity generally measures how well one thing mimics another in typical use, whereas statistical distances measure how distinguishable one thing is from another.  Perfect error-free operations have statistical distance $\delta=0$ and fidelity $F=1$, so it's common to use \textit{infidelity} $\epsilon = 1-F$ as an error rate.  Statistical distance is usually larger than infidelity, because an experiment \textit{designed} to distinguish a flawed operation from a perfect one usually detects flaws better than a ``typical'' one.

The fidelity between two pure quantum states is\footnote{Some older references define fidelity as the square root of this definition.  This definition is preferred in QCVV.} $F( \psi, \phi ) = |\braket{\psi}{\phi}|^2$.  Fidelity between mixed quantum states is
\begin{equation}
     F(\rho,\sigma) = \left(\Tr\sqrt{\sqrt{\rho}\sigma\sqrt{\rho}}\right)^2, 
\end{equation}
but in QCVV usually one state is pure and the simpler formula $F(\rho,\proj{\psi}) = \braopket{\psi}{\rho}{\psi}$ applies.  The \textit{process fidelity} between two CPTP maps $G$ and $G'$ is defined in terms of state fidelity as
\begin{equation}
    F_{\mathrm{process}}(G,G') \equiv F\Big( (G\otimes\Id)[\proj{\Psi}], (G'\otimes\Id)[\proj{\Psi}] \Big),
\end{equation}
where $\ket{\Psi}$ is a maximally entangled (``Bell'') state on two registers of the same size.  \textit{Average gate fidelity}, defined as $F_{\mathrm{avg}} = (dF_{\mathrm{process}}+1)/(d+1)$, is sometimes used instead.

The statistical distance between two quantum states, $\delta_{\mathrm{tr}}( \rho, \sigma ) = \frac12\vert \rho - \sigma \vert_1 = \frac12\Tr\sqrt{(\rho-\sigma)^2}$ is called \textit{trace distance} and derived from \textit{total variation distance} (TVD) between distributions.  The statistical distance between CPTP maps, known as \textit{diamond norm distance} is defined as an optimization (with no general closed form):
\begin{equation}
    \delta_{\diamond}(G,G') = \max_{\rho_{AB}} \delta_{\mathrm{tr}} \left( (G_A\otimes\Id_B)[\rho_{AB}], (G'_A\otimes\Id_B)[\rho_{AB}] \right).
\end{equation}
The fact that $\delta_{\diamond}$ can only be defined using an auxiliary system $B$, because entanglement with such an auxiliary system actually increases the distinguishability between $G$ and $G'$ in some cases, is a celebrated and nontrivial peculiarity of quantum information.

\section{Tomography}

Quantum tomography protocols stem from a simple and obvious idea:  since quantum states and quantum operations are described by matrices, we can figure out \textit{which} matrix it is by measuring matrix elements experimentally.  Quantum state tomography was proposed in the 1960s \cite{Gale1968-mk}, but its first experimental use was in 1989, to reconstruct the quantum state of light -- specifically of an optical mode in a reflective cavity \cite{Vogel1989-vt}.  Although optical states can be described by density matrices, they are more commonly represented by \textit{Wigner functions}, a linear representation of the mode's density matrix as a quasiprobability distribution (like a probability distribution, but not always positive) on a 2-dimensional phase space \cite{tatarskiui1983wigner}.  While a Wigner function can't be observed directly, its \textit{marginal} along any axis in phase space defines an observable probability distribution.  The earliest examples of quantum state tomography estimated these distributions along many axes and reconstructed the whole Wigner function using the Radon transform \cite{easton1984application}.  The name ``tomography'' comes from its similarity to the computer assisted tomography (CAT) scans used in medicine.

Quantum state tomography requires measuring many samples or copies of an unknown state $\rho$ -- i.e., identically prepared systems.  These are often prepared sequentially using the same system, e.g., by trapping a single atomic ion, then preparing and measuring it millions of times.  If the samples are not all identical, tomography measures their average density matrix.  The $N$ samples are divided into $M$ subsets labeled $1\ldots M$, and all $N/M$ samples in subset $m$ are measured the same way.  The outcome probabilities of each measurement are estimated from the statistics of the observed outcomes.  Then, all $M$ sets of estimated outcome probabilities are combined to estimate the density matrix $\rho$.

This works because Born's rule -- which states that the probability of any measurement outcome described by the positive operator $E_k$ is given by $\mathrm{Pr}(E_k|\rho) = \Tr(E_k\rho)$ -- is linear.  If we think of both $\rho$ and $E_k$ as vectors in the $d^2$-dimensional vector space of $d \times d$ matrices, then Born's rule is just a dot product.  Inspired by Dirac's bra-ket notation, scientists often write these vectors as $\sket{\rho}$ and $\sket{E_k}$, so that Born's rule is simply
\begin{equation}
    \mathrm{Pr}(E_k|\rho) = \sbraket{E_k}{\rho}.
\end{equation}
If a set of $d^2$ matrices $E_k$ can be found so that (1) each $E_k$ describes an outcome of some POVM measurement and (2) the vectors $\sket{E_k}$ span the space of $d\times d$ matrices, then tomography is possible.  Their probabilities define $\rho$ uniquely.  We can arrange the estimated probabilities $p_k \approx \mathrm{Pr}(E_k|\rho)$ into a vector $\vec{p}$, stack the vectors $\sbra{E_k}$ into a matrix $\mathcal{M}$, write
\begin{equation}
    \vec{p} = \mathcal{M}\sket{\rho},
\end{equation}
and invert this equation to get $\sket{\rho} = \mathcal{M}^{-1}\vec{p}$.  This is called \textit{linear inversion} tomography, and can easily be generalized to infer $\rho$ from more than $d^2$ measured probabilities by using the Moore-Penrose pseudoinverse $\mathcal{M}^+$.

CPTP maps can also be reconstructed tomographically.  This is \textit{quantum process tomography} \cite{Chuang1997-vf}.  To do process tomography, the experimenter (1) prepares many copies of several different known states $\rho_j$, (2) applies an unknown process $G$ to all of them, and (3) performs state tomography on each batch.  From the resulting data, all the probabilities
\begin{equation}
    p_{j,k} = \mathrm{Pr}(E_k|\rho_j) = \sbraopket{E_k}{G}{\rho_j}
\end{equation}
can be estimated.  If both the sets $\{\sket{E_k}\}$ and $\{\sket{\rho_j}\}$ span the space of $d\times d$ matrices, then these probabilities are sufficient to estimate the unknown CPTP map describing $G$.  Unsurprisingly, POVMs describing quantum measurements can be estimated very similarly using \textit{quantum measurement tomography} \cite{Luis1999-ek,Lundeen2009-qs}.

Quantum tomography isn't perfect, because of \textit{finite sample errors} (a.k.a.~shot noise).  For the same reason that flipping a fair coin 100 times doesn't guarantee exactly 50 heads, estimates of $\mathrm{Pr}(E_k|\rho_j)$ from $N$ samples will fluctuate around the truth by $O(1/\sqrt{N})$.  This is surprisingly consequential, because those fluctuations can cause estimated density matrices to violate the \textit{positivity constraint} $\rho \geq 0$.  (In process tomography, the estimated linear map can violate complete positivity).  Such an estimate is obviously wrong, because if taken seriously it would predict \textit{negative} probability for some as-yet-unobserved event.  This problem is routinely solved using techniques from statistical inference, mostly \textit{maximum likelihood estimation} (MLE) \cite{Derka1996-ya} and \textit{Bayesian inference} \cite{Derka1996-ya,Blume-Kohout2010-vv}.  Both MLE and Bayesian state tomography use the statistical concept of a \textit{likelihood function} $\cL(\rho) \equiv \mathrm{Pr}(\mathrm{data}_{\mathrm{observed}}|\rho)$, which is defined only on valid density matrices satisfying $\rho \geq 0$.  MLE uses numerical optimization to find and report the $\rho$ that maximizes $\cL$, while Bayesian inference starts with a prior distribution over $\rho$, multiplies it by $\cL(\rho)$ to get a posterior distribution, and uses numerical integration to find and report the posterior's mean.  MLE and Bayesian inference can be applied straightforwardly to process tomography \cite{Fiurasek2001-mp,granade2016practical}.

Large quantum registers have large Hilbert spaces, which make tomography infeasible.  An $n$-qubit register has a $2^n$-dimensional Hilbert space, so density matrices have $2^n\times 2^n = 4^n$ matrix elements to be measured, while CPTP maps have approximately $16^n$.  The practical limit is about 8-10 qubits for state tomography \cite{Haffner2005-ys,song201710}, and half that for process tomography.  However, a range of clever \textit{ans\"atze} have been proposed and used to simplify tomography, reduce its experimental cost, and extend it to larger registers.  They only work if the unknown state or process has a particular efficiently-describable form, but most ansatz tomography protocols are self-checking and can detect that their ansatz does not hold.  Prominent examples include low-rank (compressed sensing) tomography \cite{Gross2010-ck,Guta2012-nx,Flammia2012-sw}, matrix product state tomography \cite{Cramer2010-vo,Landon-Cardinal2012-nh,Ohliger2013-sp}, and neural network tomography \cite{Kieferova2017-oi,Torlai2018-wt}.

Tomography is sometimes also used to estimate or reconstruct the Hamiltonian \cite{Cole2005-aq,Shabani2011-ja} or Lindbladian \cite{Boulant2003-uu,Howard2006-uv,Samach2022-hr} that generates a quantum system's dynamics.  Generators of dynamical evolutions can be sparse in ways that density matrices or process matrices usually are not, which makes it possible to estimate them efficiently using compressed sensing techniques \cite{donoho2006compressed}.  Learning generators can also provide more useful predictive power than process tomography, because it can enable scientists to design good control operations, by revealing not just what a register is doing, but how it is likely to respond to new influences.

\section{Randomized benchmarking}

Randomized benchmarking (RB) protocols were invented in the mid-to-late 2000s \cite{Emerson2005-sd,Knill2008-qe} to probe quantum gate performance.  Quantum process tomography presented two critical problems: it becomes unreliable in the presence of state preparation and measurement (SPAM) errors; and its complexity grows as $O(16^n)$ for $n$-qubit systems.  RB avoids both, but at the price of revealing far less.  In its original form, RB measures just \textit{one} number -- an average error rate -- to describe an entire set of quantum gates.

The standard RB protocol \cite{Magesan2011-ra} measures an average error rate for a set of logic gates $\mathbb{G}_n$ on $n\geq1$ qubits that is (1) a group and (2) a unitary 2-design.  The most common example is the $n$-qubit Clifford group $\mathbb{C}_n$, and most RB experiments benchmark 1- or 2-qubit Clifford gates, but any set of gates forming a group representation that acts irreducibly on the traceless subspace of $d\times d$ matrices meets the conditions.  Standard RB measures an average error rate called \textit{average gate set infidelity}, which under most circumstances is approximately equal to the average (over all gates) of the average gate infidelity between each gate and its ideal target \cite{Proctor2017-ru,Wallman2018-wy}.  It does so via the following algorithm \cite{Magesan2011-ra}:
\begin{enumerate}
    \item Choose many uniformly random sequences of gates, of varying lengths.
    \item Transform each sequence into a \textit{motion-reversal} (identity) sequence by appending a unique ``inversion gate'', chosen using group theory, to perform the inverse of the group element that the original sequence implemented.
    \item Execute each sequence on the quantum processor, following it with a measurement that checks whether the processor was successfully returned to its initial state.  
    \item Plot, versus sequence length $m$, the fraction $F(m)$ of the runs that successfully returned to the initial state.
    \item Fit an exponential decay $F(m) = Ap^m + B$ to the data, and report the average error rate of the gates as $r = (d-1)(1-p)/d$, where $d$ is the dimension of the register's Hilbert space.
\end{enumerate}
The mathematical theory of RB is surprisingly deep, but standard analyses rely on the concept of \textit{group twirling}.  If each noisy gate can be described as its ideal counterpart followed by an \textit{error process} that is close to the identity process \textit{and} independent of which gate was performed, then Schur's Lemma can be used to show that the error process will be ``twirled'' like a block on a lathe, so that its effects display the same symmetry as the gate set.  Because the gate set was chosen to be a unitary 2-design, it is \textit{extremely} symmetric, and the effective noise channel (averaged over random sequences) is uniformly depolarizing.  As a result, a single uniform exponential decay in $F(m)$ will be observed, with only the $A$ and $B$ coefficients depending at all on SPAM errors.

The original ``Clifford RB'' protocol described above has been generalized extensively.  Other unitary 2-designs besides the Clifford group can be used.  Native-gate or ``direct'' RB protocols \cite{Proctor2019-ma} remove the need for a full group, and quantify the average error rate of a set of gates that merely \textit{generate} a unitary 2-design.  Character RB \cite{Helsen2019-nx} and the more general \textit{filtered RB} techniques \cite{helsen2022general} benchmark gate sets that generate reducible groups, and extract more than 1 error parameter in the process.

Further generalizations build on the RB foundation to measure quantities and error rates \textit{beyond} average gate error.  Interleaved RB \cite{magesan2012efficient} estimates error rates of individual gates, albeit with some critical limitations.  Simultaneous RB \cite{gambetta2012characterization} measures error rates of individual subsystems (e.g. single qubits) in parallel, inclusive of crosstalk errors.  Purity benchmarking \cite{wallman2015estimating} separates the contributions of coherent and incoherent (stochastic) errors to the RB error rate.  Cycle benchmarking \cite{Erhard2019-ig} combines techniques from character RB and interleaved RB to efficiently estimate the rates of many different classes of error in an $n$-qubit circuit layer comprising many gates in parallel.  Universal-gate RB \cite{Hines2023-ph} and cross-entropy benchmarking \cite{Boixo2018-wy} measure the average error rate of computationally universal gate sets that don't generate a closed discrete group at all. RB protocols have been proposed and used to quantify leakage errors \cite{chasseur2015complete} and errors in mid-circuit measurements \cite{hothem2024measuring,govia2023randomized} used for quantum error correction.  Finally, \textit{scalable} RB protocols \cite{Proctor2022-nt,hines2024fully} use a range of tricks to reduce the overhead of benchmarking the performance of $n$ qubits at once.

\section{Holistic benchmarks}

The appearance of integrated quantum processors with 5+ qubits (circa 2016) created demand for techniques to benchmark their performance \cite{Linke2017-ht}, and enthusiasm for standardized suites of benchmarks like those used for classical computing \cite{dongarra2003linpack,henning2006spec}. These ``holistic benchmarks'' measure one or more quality metrics that are intended to concisely summarize an integrated quantum computer's performance, enabling comparisons between different systems and the tracking of technological progress \cite{proctor2025benchmarking}. Holistic benchmarks are typically designed to measure heuristic proxies for overall system quality.  In contrast to the model-focused and less scalable approach of characterization protocols like tomography, they are usually intended to be executable on registers of tens, hundreds, or even thousands of qubits.

The earliest holistic benchmarks consisted of simply running tiny instances of quantum algorithms and measuring their success probabilities or process fidelities \cite{vandersypen2001experimental, Lucero2012-kd, Linke2017-ht}. Over the last five years, this approach has been formalized in a variety of algorithm- and application-focused benchmarking suites \cite{Tomesh2022-nu, Lubinski2023-zy, Li2020-ry, Quetschlich2023-bg, Martiel2021-vp, Chatterjee2024-py} that prescribe running quantum algorithms (e.g.~phase estimation) with various problem instances, or algorithmic primitives (e.g.~the quantum Fourier transform). Many of these benchmarks are designed to quantify the joint performance of the the entire quantum computing ``stack'', which includes many classical aspects like compilers, routers, or error mitigation in addition to the quantum register itself. But extracting objective cross-platform metrics from algorithmic benchmarks is tricky, in part because present-day quantum computers are too small and noisy to run ``utility-scale'' quantum algorithms \cite{proctor2025benchmarking}. The relative performance of two quantum computers at solving a tiny problem or executing a small subroutine (e.g., a 10-qubit quantum Fourier transform) does not \emph{a priori} predict their future value.

Quantum computers can also be benchmarked by how well they run quantum circuits that are specifically designed for benchmarking, and do not implement a recognized quantum algorithm. This is the approach taken by two of the best-known holistic benchmarks, IBM's \textit{quantum volume} benchmark \cite{Cross2019-el} and Google Quantum AI's \textit{cross-entropy benchmarking} \cite{Boixo2018-wy}. Both were developed in the 2010s, and involve running $n$-qubit random circuits with $d$ layers. Those circuits' structure, and the analysis of the resulting data, are specific to each method. The quantum volume benchmark specifies ``square'' random circuits with $n=d$, and the quantum computer's quantum volume is defined as $\textrm{QV} = 2^{n_{\max}}$ where $n_{\max}$ is the size of the largest circuits whose success rate (as quantified by heavy output probability \cite{Cross2019-el}) exceeds a somewhat arbitrary threshold of $2/3$. Quantum volume is the paradigmatic \emph{full-stack benchmark}: it permits almost arbitrary compilation of its circuits, and therefore it jointly tests a quantum computer's qubits and compilers. Both quantum volume and cross-entropy benchmarking have significant scalability limitations, e.g., because they require classical simulations of very general $n$-qubit circuits, but they have both been adapted into scalable techniques \cite{Hines2023-ph, Hines2024-ae, Chen2022-hd}. 

One purpose of holistic benchmarks is to quantify and track technological progress towards useful quantum computation.  This goal motivates \emph{capability benchmarks}, which attempt to estimate and describe the set of circuits that a quantum computer can run with high success rate \cite{Hothem2024-rc}.  A given quantum computer's capability can be communicated using a \emph{capability region} that predicts whether a circuit can (or cannot) be run successfully from summary features of that circuit \cite{Proctor2021-wt, proctor2025benchmarking}. The most well-known kind of capability benchmarking is \emph{volumetric benchmarking} \cite{Blume-Kohout2020-de, Proctor2021-wt}, which extends the concept of quantum volume to map out a quantum computer's ability to run circuits of varying widths and depths. Capability benchmarking can be performed with almost any class of quantum circuit, and it can be applied to any number of qubits by using techniques for scalable assessment of circuit fidelities \cite{Proctor2022-zs, Seth2025-zz}.  

\section{Advanced topics and future directions}
Quantum characterization, verification, and validation is a rapidly evolving field with many advanced topics for further exploration. This section reviews a few of the most active areas of ongoing research, and highlights potential future directions in QCVV.

\vspace{0.3cm}
\noindent\textit{Gate set tomography}:  State and process tomography assume the existence of a perfectly calibrated set of state preparation and/or measurement operations, but errors in these ``SPAM'' operations can bias the results of tomography.  \textit{Gate set tomography} (GST) was developed around 2012-13 to provide a self-calibrating or calibration-free framework for characterizing quantum operations \cite{Merkel2013-me,Blume-Kohout2013-mf,Nielsen2021-ck}.  GST uses data from a large ensemble of quantum circuits to reconstruct a mathematical model for the entire gate set. This model can be used to predict other circuits, or analyzed to yield performance metrics and error models for each gate, state preparation, or measurement operation in the gate set. GST can distinguish and quantify coherent and incoherent errors in individual gates, relational errors between them, and standard error metrics such as fidelity or diamond-norm error \cite{Aharonov1998-jc}. 

GST spurred the realization that gate sets have a \textit{gauge freedom}.  Many seemingly distinct mathematical models for quantum gate sets produce identical observable behavior in quantum circuits \cite{Nielsen2021-ck}. The presence of this gauge freedom in gate set models remains a stumbling block for standard approaches to QCVV. In practice, it can be addressed through a gauge fixing procedure that selects a representation of the gate set that closely approximates the target gate set, according to a suitable norm.

GST has emerged as an alternative to RB for characterizing small quantum systems (1-2 qubits). It offers the advantage of providing richer information about gate performance and the ability to construct predictive models of quantum operations. However, GST is more experimentally demanding and can be more sensitive to non-Markovian effects.

\vspace{0.3cm}
\noindent\textit{Phase estimation}:  Although the matrices describing quantum operations can vary significantly depending on the gauge choice, their eigenvalues are gauge-invariant. Phase estimation offers a method to measure the complex arguments (the \textit{phase}) of these eigenvalues, making it a critical component of many quantum algorithms, particularly those related to quantum simulation. 

In the context of QCVV, phase estimation can be employed to quantify coherent (unitary) errors in quantum gates. The simplest quantum phase estimation protocols are based on Ramsey spectroscopy, a technique that has been in continuous use since its development in 1949 \cite{Ramsey1950-od}. However, traditional phase estimation methods are susceptible to SPAM errors, which can result in biased estimates.  \emph{Robust phase estimation} (RPE) \cite{Kimmel2015-tj} is resilient against many forms of SPAM error, and can estimate individual gates' phases with high accuracy using straightforward data analysis to mitigate state preparation errors.

\vspace{0.3cm}
\noindent\textit{Pauli noise learning}:  
A generic noise channel on $n$ qubits requires approximately $16^n$ numbers to specify. But under certain conditions, such as within quantum error correction circuits or under twirling with the Pauli group, the qubit dynamics can be captured completely in terms of the $4^n$ parameters of an $n$-qubit Pauli channel. In a Pauli channel, the errors behave as though random Pauli operations (phase-flips, bit-flips, etc.) have been inserted into the circuit, each at its own rate.  Pauli noise learning focuses on learning some or all of these error rates. Techniques for Pauli noise learning include cycle error reconstruction \cite{Carignan-Dugas2023-uk}, average circuit eigenvalue sampling (ACES) \cite{Flammia2022-ij}, and compressed sensing approaches \cite{harper2021fast}. 

Knowledge of effective Pauli channels can be extremely useful for identifying high-weight errors, routing and compiling quantum circuits \cite{Wagner2025-db,Murali2019-tp}, understanding the performance of quantum error correction \cite{harper2023learning}, and optimizing decoders for quantum error correction \cite{Wagner2022-gb}. 

\vspace{0.3cm}
\noindent\textit{Characterization of mid-circuit measurements}: 
Mid-circuit measurements (MCMs) are measurements that are performed during the execution of a quantum algorithm rather than at its end.  Because they must leave the measured subsystem (e.g.~qubit) in a predictable computational state rather than destroying it, MCMs are sometimes called \textit{nondemolition measurements} \cite{ralph2006quantum}. They are essential for implementing quantum error correction. QCVV on MCM operations has received greater attention as quantum error correction demonstrations have become increasingly feasible.

Simple models for errors in MCMs quantify the probability of incorrect readout, or the fidelity of the post-measurement quantum state.  \emph{Quantum instruments} \cite{Davies1970-oe}, a mathematical model for MCMs, provide a comprehensive description of Markovian errors in MCMs.  They combine the features of a positive operator-valued measure or POVM (the usual model for circuit-terminating measurements) with those of a quantum process (the usual model for noisy gates).  There are QCVV protocols for tomographic reconstruction of the quantum instrument describing a MCM \cite{Blumoff2016-uk,Wagner2020-bt,Rudinger2022-at}.  MCMs appearing only in Pauli-twirled circuits can be characterized using techniques from Pauli noise learning \cite{hines2025pauli,zhang2025generalized}. Finally, RB variants can be used to assess the total error present in an MCM \cite{hothem2024measuring}. 

\vspace{0.3cm}
\noindent\textit{Non-Markovian dynamics and noise spectroscopy}: 
When a quantum register is influenced by an environment with nontrivial temporal correlations, the resulting register dynamics will be non-Markovian. There are many ways for a system to display non-Markovianity, ranging from relatively benign parameter instabilities to difficult-to-diagnose bugs in control systems. Some common examples of non-Markovian noise sources include: parameter drifts in classical control fields \cite{Proctor2019-oi}, persistent quantum spin baths \cite{Prokof-ev2000-oi}, presence of leakage states \cite{Wallman2016-xe,Wood2018-wi}, resonator-induced distortion of control pulses, and $1/f$ environmental noise \cite{Schriefl2006-vp}. 

Temporally correlated noise sources can present a significant challenge for traditional QCVV techniques, which typically assume the dominant errors to be Markovian. This assumption simplifies the models, experiments, and analyses, but can lead to biased or nonsensical estimates when it is violated \cite{Figueroa-Romero2021-ym,Nielsen2021-ck}. Extending QCVV techniques to enable characterization of non-Markovian dynamics is an active area of research.  It includes the development of metrics that quantify non-Markovianity \cite{Breuer2012-il}, construction of models that capture non-Markovian effects \cite{Rivas2014-cz}, and design of experiments that are sensitive to non-Markovian model parameters \cite{White2020-de}. 

Since the early 2000's, a key focus of research into non-Markovian noise characterization has been \emph{noise spectroscopy}, a family of techniques for learning the power spectral density of environmental noise acting on quantum systems \cite{Wu2013-nz,Paz-Silva2017-le}. More recently, process tensors have emerged as a framework for modeling and characterizing quantum systems that are coupled to persistent quantum environments \cite{White2020-de}. Additionally, several traditional QCVV protocols have been extended to capture non-Markovian dynamics, including GST \cite{Proctor2019-oi,Li2024-gg} and RB \cite{Figueroa-Romero2021-ym}. 

\vspace{0.3cm}
\noindent\textit{Randomized measurements and classical shadows}: Learning a complete description of a general $n$-qubit quantum state or process is exponentially expensive in $n$, but a partial description is sufficient for many purposes and can often be learned much more efficiently \cite{Elben2022-bz, Aaronson2019-yg, Aaronson2018-nh, Paini2019-fd, Huang2020-pu}. Many techniques for learning partial representations of a quantum state $\rho$ are based on \emph{randomized measurements}. If a sequence of randomly chosen measurements is performed on samples of $\rho$, the list of their outcomes defines a \emph{classical shadow} of $\rho$ \cite{Paini2019-fd, Huang2020-pu}. A classical shadow can be used to accurately predict many (but not all) observables of $\rho$ from many fewer measurements than needed for state tomography.

Randomized measurement and classical shadow methods have a variety of applications outside of QCVV \cite{Elben2022-bz}. Within QCVV, the most well-known application is \emph{direct fidelity estimation} (DFE) \cite{Flammia2011-qj, Moussa2012-rq}. DFE is a method for estimating the fidelity $F(\rho,\psi)$ between an actual state $\rho$ and some known ideal state $\psi$, by measuring $\rho$ in random Pauli bases. DFE is more efficient than the brute-force approach of estimating $F(\rho,\psi)$ by performing state tomography on $\rho$, but it is still exponentially costly for general $\psi$. However, DFE is efficient for certain families of states. Most notably, a \textit{constant} number of measurements is sufficient to estimate $F(\rho,\psi)$ when $\psi$ is a stabilizer state (a state created by applying Clifford gates to qubits initialized in the computational basis). DFE can also be applied to estimating the fidelity of processes, and it is a primitive in a variety of QCVV methods \cite{Erhard2019-ig, hines2024fully, Seth2025-zz}.

\vspace{0.3cm}
\noindent\textit{Verification and Certification}:  Another way to frame the question ``How good is a quantum device?'' is as \textit{certification} or \textit{verification} that it works correctly.  Certification means testing the hypothesis that the device works correctly to within some margin \cite{KlieschPRXQ21}.  In practice \cite{Eisert2020-xd}, certification often boils down to using a QCVV protocol to measure a quality parameter (with confidence bounds) and then certifying the device if the quality parameter is above a threshold.  However, this approach has motivated some specific protocols, such as blind quantum computing \cite{fitzsimons2017private}, that focus on rigorous verification of quantum computational functionality.

\vspace{0.3cm}
\noindent\textit{Performance of logical qubits and quantum error correction primitives}: Increasingly, the road to utility-scale quantum computing appears to go through error-corrected \textit{logical qubits} constructed from multiple physical qubits and designed to be fault tolerant with (eventually) very low error rates.  Accordingly, the focus of QCVV is shifting towards the characterization and benchmarking of logical qubits.  This poses new challenges, because the nature and behavior of logical qubits is qualitatively different from that of physical qubits \cite{eastin2009restrictions}.

One aspect of logical-qubit QCVV is characterizing and modeling the \textit{components} of logical qubits and quantum error correction primitives, such as few-qubit parity checks \cite{Chow2014-bc,Takita2016-po}.  Logical qubit systems have large Hilbert spaces that make direct simulation of their behavior prohibitively expensive, motivating effective models that describe that behavior accurately but efficiently.  As better logical qubits are demonstrated in experiments, QCVV protocols for directly benchmarking their performance have emerged.  Logical error rate, or the rate of its decrease with code distance \cite{Google2025-xr} is an obvious metric.  However, some recent demonstrations have gone further and used the data produced by QEC to characterize and benchmark the logical qubit's internal behavior \cite{Google2021-gu}.

\section{Further reading}

QCVV serves the needs of quantum information and quantum computing, and a good general understanding of that field is a prerequisite for understanding QCVV.  Arguably the best introduction to the field is \textit{Quantum Computation and Quantum Information}, by Michael A.~Nielsen and Isaac L.~Chuang \cite{NielsenBook2000}.  Chapter 8 is especially useful, as are chapters 2, 4, and 9.

Concepts and methods from statistics and statistical inference are ubiquitous in QCVV.  \textit{All of Statistics}, by Larry Wasserman \cite{wassermanBook2013}, is an accessible gateway to this field.  Chapters 6 and 9 are especially relevant to quantum tomography.

There are, as of this writing, no textbooks on QCVV that we recommend.  However, a comprehensive tutorial paper by Akel Hashim \textit{et al.~} \cite{hashim2024practical} provides a reliable introduction to all the concepts discussed here and more.  

There are also several review articles or books addressing QCVV as a whole \cite{Eisert2020-xd}, or specific topics within it including tomography \cite{paris2004quantum}, certification \cite{KlieschPRXQ21} and verification \cite{Yu2022-tu,Gheorghiu2019-fu}, self-testing \cite{Supic2020-vw}, learning \cite{Gebhart2023-dr}, non-Markovian phenomena \cite{Milz2021-ad}, and benchmarking \cite{proctor2025benchmarking}.  A set of lecture notes by Martin Kliesch \cite{Kliesch2019-ak} offers a pedagogical approach to a subset of QCVV.

%------------- Acknowledgements
\section{Acknowledgements}
This material was funded in part by the U.S. Department of Energy, Office of Science, Office of Advanced Scientific Computing Research, Quantum Testbed Pathfinder Program. Sandia National Laboratories is a multi-program laboratory managed and operated by National Technology and Engineering Solutions of Sandia, LLC., a wholly owned subsidiary of Honeywell International, Inc., for the U.S. Department of Energy's National Nuclear Security Administration under contract DE-NA-0003525. All statements of fact, opinion or conclusions contained herein are those of the authors and should not be construed as representing the official views or policies of the U.S. Department of Energy or the U.S. Government.

%------------- End matter
\bibliography{bibliography.bib}
\end{document}